\documentclass[aps,prl,showpacs,amssymb]{revtex4}
\usepackage{graphicx}% Include figure files
\usepackage{bm}% bold math

\def\ve {\varepsilon}
\def\ep {\epsilon}

\def\e2 {\epsilon-\epsilon_k}
\def\be {\begin{equation}}
\def\ee {\end{equation}}
\def\bea {\begin{eqnarray}}
\def\eea {\end{eqnarray}}
\def\om {\omega}

\def\cd {c^{\dagger}}
\def\dd {d^{\dagger}}
\def\ua {\uparrow}
\def\da {\downarrow}
\def\si {\sigma}

\begin{document}

\centerline{European Physical Journal B {\bf 73}, 483-487 (2010)}

\title{Microscopic Bardeen-Cooper-Schrieffer formulation of the critical
temperature of multilayer copper-oxide superconductors}

\author{ George Kastrinakis$^*$}

\affiliation{ Institute of Electronic Structure and Laser (IESL), 
Foundation for Research and Technology - Hellas (FORTH),
P.O. Box 1527, Iraklio, Crete 71110, Greece}

\date{Received 22 October 2009 / In final form 8 December 2009}

\begin{abstract}

We study superconductivity in multilayer copper oxides, in the frame
of a realistic microscopic formulation. 
Solving the full temperature dependent BCS gap equations,
we obtain a maximum in the 
transition temperature $T_c$ for M=3 or 4 CuO$_2$ layers in the unit cell
for appropriate values of the interlayer tunneling (negative pair
tunneling), and via the consideration of the doping 
imbalance between the inner and outer layers. 
This is 
the ubiquitous experimental result for Ca intercalated copper oxides,
as opposed to other intercalating elements. 
Further, using a restricted set of parameters, we obtain an exact
fit of $T_c$(M=1-4) for five different Ca intercalated
homologuous copper oxide families.

\end{abstract}


\maketitle

A long standing puzzle of the high transition temperature $T_c$
cuprate superconductors is the dependence of $T_c$ on the number 
M of CuO$_2$ layers in the unit cell. Experiments show ubiquously that 
within any Ca intercalated homologuous cuprate family, i.e. for materials 
sharing the same charge reservoir block but having varying M (here
Ca is between the layers),
$T_c$ is maximum for M=3 - a summary of experiments appears in e.g. 
\cite{lin} - 
with the M=4 exception of the 
Tl$_2$Ba$_2$Ca$_{M-1}$Cu$_M$O$_{2M+3+\delta}$ family \cite{burn}, 
following a dome-type dependence on M. 
That is at optimum doping for any M. As mentioned in \cite{le}, other
spacing elements, such as Y, Ba and Sr yield lower $T_c$ for the 
multilayer materials compared to the single layer material. 
Setting aside the central issue of superconductivity within a
single CuO$_2$ layer, the issue of what determines $T_c$ in
a multilayer system is a very important one. Inter alia, it
points to the role of out-of-plane mechanisms and to how $T_c$ can be 
maximized for any given cuprate family.

This issue has been treated in a number of theoretical papers, as in
\cite{tes,sta,ch,spa,liec,car,lin,cha}, with limited success.
E.g. in ref. \cite{lin} a Bardeen-Cooper-Schrieffer (BCS) model 
(different from ours) was considered,
and a saturating increase of $T_c$ as a function of M=1-$\infty$ was obtained.
In ref. \cite{cha} the dome with a maximum for M=3 was obtained for 
the superconducting order parameter (not directly for $T_c$ itself) in the 
frame of a Landau-Ginzburg functional treatment, and by appealing to the 
putative ddw order parameter.

The cuprates fall into the realm of strong correlations, thus calling for an
Eliashberg-type treatment \cite{scala,gk1,gk2}. Given the lack of a definitive
theory for the single CuO$_2$ layer superconducting state, our BCS
treatment is a decent attempt towards the understanding of the multilayer 
cuprates.
We hereby consider a realistic microscopic model in the frame of the
BCS formulation \cite{dg}, which does yield the experimental fact
of the maximum $T_c$ for M=3 or 4 for a range of the parameters considered.
That is for moderately small single particle and pair tunneling between
successive CuO$_2$ layers. As experiments indicate that $T_c$ decreases
steadily for M$>$3 (with the aforementioned exception), we content ourselves 
with calculating $T_c$ for the 
cases M=1-4, assuming that the trends found also apply for M$>$4.
The complex calculations required for M$>$4 are out of the scope of the
present paper.

We consider the BCS-type Hamiltonian for M CuO$_2$ layers in the unit cell
\bea
H = \sum_{i;k,\si} \varepsilon_{i,k} \; \dd_{i,k,\si} d_{i,k,\si}
+ \sum_{i;k,p}V_i(k,p)\; \dd_{i,k,\ua} \dd_{i,-k,\da} d_{i,-p,\da}d_{i,p,\ua}
\\ \nonumber
+\sum_{<il>,k,\si} t_{\perp,k} \; \dd_{i,k,\si}d_{l,k,\si} 
+\sum_{<il>,k} T_{k} \; \dd_{i,k,\ua} \dd_{i,-k,\da} 
d_{l,-k,\da} d_{l,k,\ua} 
 \;\;.
\eea
The creation/annihilation operators $\dd_{i,k,\si}/d_{i,k,\si}$ describe 
electrons in the $i$-th CuO$_2$ layer in the unit cell, interacting via
$V_i(k,p)$, and $i=$1-M. $\varepsilon_{i,k} = \ep_{i,k}-\mu_i$, with
the dispersion $ \ep_{i,k} = -2 t_i (\cos{k_x}+\cos{k_y}) 
- 4 t_i'\cos{k_x} \cos{k_y} - 2 t_i''(\cos{2k_x}+\cos{2k_y}) $, 
$k_x,k_y=[-\pi,\pi]$, and the 
chemical potential $\mu_i$ of the $i$-th
layer. The (in-plane momentum conserving) coupling between successive neighbor 
CuO$_2$ layers $<il>$ for single electron tunneling is \cite{lie2}
\be
t_{\perp,k} = t_o ( \cos{k_x} - \cos{k_y} )^2 \;\;,
\ee
while for pair tunneling is (proportional to $t_{\perp,k}^2$, 
e.g. c.f.\cite{ch} and below) 
\be
T_k = T_o ( \cos{k_x} - \cos{k_y} )^4 \;\;.
\ee

We diagonalize the single particle kinetic energy part of the Hamiltonian.
The original operators $d_{i,k,\si}$ are given in terms of the new ones
$c_{i,k,\si}$
as $\vec{d}_M = U_M \vec{c}_M$, 
$\vec{d}_M = (d_{1,k,\si}, d_{2,k,\si}, ...,d_{M,k,\si})$ \cite{oper}. 
In the new basis, we consider variational BCS states of the type
\be
|\Psi> = \prod_{i,k} (u_{i,k} + v_{i,k} \; \cd_{i,k,\ua} \cd_{i,-k,\da})|0>
\;\; ,
\ee
with $u_{i,k},v_{i,k}$ the well known coherence factors. 

For M=2 layers $c_{1,2}=(d_1\pm d_2)/\sqrt{2}$, and the transformed 
Hamiltonian is
\bea
H = \sum_{i;k,\si} \xi_{i,k} \; \cd_{i,k,\si} c_{i,k,\si} + \frac{1}{2}
\sum_{k,p} V(k,p) [ \cd_{1,k,\ua} \cd_{1,-k,\da} c_{1,-p,\da} c_{1,p,\ua}
+ \cd_{1,k,\ua} \cd_{1,-k,\da} c_{2,-p,\da}c_{2,p,\ua} + 
(1\leftrightarrow 2)] \nonumber \\
+ \frac{1}{2} \sum_k T_k [\cd_{1,k,\ua} \cd_{1,-k,\da} c_{1,-k,\da} c_{1,k,\ua}
+ \cd_{1,k,\ua} \cd_{1,-k,\da} c_{2,-k,\da}c_{2,k,\ua} + 
(1\leftrightarrow 2)] 
\;\;.
\eea
Here $\xi_{1,k} = \varepsilon_k - t_{\perp,k}$, 
$\xi_{2,k} = \varepsilon_k + t_{\perp,k}$ correspond to the bonding and
antibonding states. The two initial layers are equivalent, and hence 
$V_1=V_2=V$. We also made use of the fact that the only non-zero matrix 
elements with four operators $c_{i,k,\si}$ ($<O>=<\Psi|O|\Psi>$) are
- c.f. ref. \cite{dg} and below for $u_{i,k},v_{i,k}$ and $f_{i,k}$
\bea
<\cd_{i,k,\ua} \cd_{i,-k,\da} c_{j,-p,\da}c_{j,p,\ua}>
=v_{i,k}u_{i,k}v_{j,p}u_{j,p} (1-2 f_{i,k})(1-2 f_{j,p})\;\; .
\eea

Bearing in mind that
%\be
$u_{i,k}^2+v_{i,k}^2=1, \;\; 0\leq u_{i,k}^2,v_{i,k}^2 \leq 1 \;\;$,
%\ee
we treat the coherence factors as $u_{i,k}=\cos(\theta_{i,k})$,
$v_{i,k}=\sin(\theta_{i,k})$. Minimizing $<H>=<\Psi|H|\Psi>$ with respect 
to $\theta_{i,k}$ yields the gap equations
\be
0= -2\; \xi_{i,k} v_{i,k} u_{i,k} +\Delta_{i,k}
[u_{i,k}^2-v_{i,k}^2] \;\;,   \label{gapeq} \\
\ee
while minimizing $<F=H-T S>$, with the entropy
$S=-2 \sum_{i,k} \left[ f_{i,k} \ln f_{i,k} +
(1-f_{i,k}) \ln (1-f_{i,k}) \right]$ and the temperature $T=1/\beta$, 
with respect to the thermal factors $f_{i,k}$ yields
\be
f_{i,k} = \frac{1}{1+e^{\beta E_{i,k}} } \;\;,\;\;  
E_{i,k} = \frac{ \xi_{i,k} }{ u_{i,k}^2-v_{i,k}^2 } \;\;. \label{fieq}
\ee
The gaps $\Delta_{i,k}$ are a sum of a diagonal (in the layer index $i$) part
$G_{i,k}$ and a non-diagonal part $g_{i,k}$.
\bea 
\Delta_{i,k} = G_{i,k} + g_{i,k}\;\; , \;\;
 G_{i,k} = -\frac{1}{2}\sum_{p} V(k,p) \; v_{i,p} u_{i,p}\; 
\tanh(\beta E_{i,p}/2) \;
-\; \frac{1}{2}T_k v_{i,k} u_{i,k}\; \tanh(\beta E_{i,k}/2)\;\;, \;\;   \\
g_{i,k} = - \frac{1}{2}\sum_{p} V(k,p) \; v_{j,p} u_{j,p}\; 
\tanh(\beta E_{j,p}/2) \;
- \; \frac{1}{2}T_k v_{j,k} u_{j,k}\; \tanh(\beta E_{j,k}/2)\;, 
\;\;
(i,j) = \{(1,2),(2,1)\}   \nonumber
 \;\;.
\eea
Here $\Delta_{1,k}=\Delta_{2,k}$.
We note that setting $i=j$ and $T_k=t_{\perp,k}=0$ reduces these equations
to the usual gap equation $0=- 2 \; \xi_{k} v_{k} u_{k} + \Delta_{k}
[u_{k}^2-v_{k}^2]$, with 
$\Delta_{k} = -\sum_{p} V(k,p) \; v_{p} u_{p} \tanh(\beta E_{p}/2)$ and
$E_{k} = \sqrt{\xi_{k}^2+\Delta_{k}^2} = \xi_{k}/( u_{k}^2- v_{k}^2)$.
But, note that we do {\em not} enforce a relation of the type
$E_{i,k} \sim \sqrt{\Delta_{i,k}^2+\xi_{i,k}^2 }$ for M$>$1.

The general form of the transformed Hamiltonian is
\be
H = \sum_{i;k,\si} \xi_{i,k} \; \cd_{i,k,\si} c_{i,k,\si}
+ \sum_{i,j;\;k,p} w_{ij}(k,p) \;
\cd_{i,k,\ua} \cd_{i,-k,\da} c_{j,-p,\da} c_{j,p,\ua} \;\;. \label{exiw}
\ee
The coefficients $w_{ji}(k,p)$ are determined below.
Thus we obtain as above the gap equations for $u_{i,k},v_{i,k},f_{i,k}$ 
in the general form (\ref{gapeq}),(\ref{fieq}), but now with
\be
G_{i,k} = - \; \sum_{p} w_{ii}(k,p) \; v_{i,p} u_{i,p} \; 
\tanh(\beta E_{i,p}/2) \;\;, \;\;
g_{i,k} = - \frac{1}{2}\sum_{j\neq i ; p} [ w_{ij}(k,p) + w_{ji}(k,p)]
\; v_{j,p} u_{j,p} \; \tanh(\beta E_{j,p}/2) \;\;.
\ee
%\vspace{.3cm}

For M=3 layers, $\xi_{1,k}=\varepsilon_{1,k}, \;
\xi_{(2,3),k}=(\varepsilon_{1,k}+\varepsilon_{2,k}\mp d)/2$, with
$d=\sqrt{a^2+8 t_{\perp,k}^2}, \; a=\ve_{1,k}-\ve_{2,k} $.
Also
\be
w_{ij}(k,p) = V_1(k,p)[ A_{ij}^{(1)}(k,p)+A_{ij}^{(3)}(k,p)]
+ V_2(k,p) A_{ij}^{(2)}(k,p) + E_{ij}(k) \; \delta_{kp}\;\;.
\ee
The terms $A_{ij}^{(m)}$ correspond to the initial 
layers $m$=1-3, with layers 1 and 3 being equivalent: 
$\varepsilon_{1,k}=\varepsilon_{3,k}$.
We have 
\bea
A_{ij}^{(1)}(k,p)=A_{ij}^{(3)}(k,p)=b_i^2(k) b_j^2(p) \;\;,\;\; 
A_{ij}^{(2)}(k,p)=g_i^2(k) g_j^2(p) \;\;,   \label{exi3}   \\
E_{ij}(k)=2 T_k \; [\; b_i^2(k) g_j^2(k) + b_j^2(k) g_i^2(k) \;]
\;\;,  \nonumber
\eea
with $b_1(k)=1/\sqrt{2},\; b_2(k)=(1-a/d)u_1/4, \;b_3(k)=(1+a/d)u_2/4,
\;g_1(k)=0, \;g_2(k)=t_{\perp,k} u_1/d, \;g_3(k)=t_{\perp,k} u_2/d$ and
$\;u_1=\sqrt{2+(a+d)^2/(2 t_{\perp,k})^2}, 
\;u_2=\sqrt{2+(a-d)^2/(2 t_{\perp,k})^2}$.

%\vspace{.3cm}

For M=4 layers
$\xi_{(1,2),k}=(\varepsilon_{1,k}+\varepsilon_{2,k} +t_{\perp,k} \mp s_{1})/2,
\;
\xi_{(3,4),k}=(\varepsilon_{1,k}+\varepsilon_{2,k} -t_{\perp,k}\mp s_{2})/2$, 
with
$d=\ve_{1,k}-\ve_{2,k}, 
\;s_{1,2}=\sqrt{d^2 \pm 2\; t_{\perp,k}\; d+5 \;t_{\perp,k}^2}$.
Here
\be
w_{ij}(k,p) = V_1(k,p)[ A_{ij}^{(1)}(k,p)+A_{ij}^{(4)}(k,p)]
+ V_2(k,p) [ A_{ij}^{(2)}(k,p)+A_{ij}^{(3)}(k,p)] 
+  E_{ij}(k) \; \delta_{kp}\;\;.
\ee
The terms $A_{ij}^{(m)}$ correspond to the initial 
layers 1-4, with layers 1 and 4 and also 2 and 3 being equivalent.
Now
\bea
A_{ij}^{(1)}(k,p)=A_{ij}^{(4)}(k,p)=b_i^2(k) b_j^2(p) \;\;,\;\; 
A_{ij}^{(2)}(k,p)=A_{ij}^{(3)}(k,p)=g_i^2(k) g_j^2(p) \;\;, \label{exi4} \\
E_{ij}(k)=2 T_k \;
[ \;b_i^2(k) g_j^2(k) + b_j^2(k) g_i^2(k) + g_i^2(k) g_j^2(k) \;]
\;\;,  \nonumber
\eea
with $b_1(k)=u_1(d+t_{\perp,k}-s_1)/(4 s_1), 
\; b_2(k)=u_2(d+t_{\perp,k}+s_1)/(4 s_1), 
\;b_3(k)=u_3(-d+t_{\perp,k}+s_2)/(4 s_2),
\;b_4(k)=u_4(d-t_{\perp,k}+s_2)/(4 s_2), \;g_1(k)=u_1 t_{\perp,k}/(2 s_1),
\;g_2(k)=u_2 t_{\perp,k}/(2 s_1), \;g_3(k)=u_3 t_{\perp,k}/(2 s_2),
\;g_4(k)=u_4 t_{\perp,k}/(2 s_2)$ and
$u_1=\sqrt{2+(d+t_{\perp,k}+s_1)^2/(2 t_{\perp,k}^2)}, 
\;u_2=\sqrt{2+(d+t_{\perp,k}-s_1)^2/(2 t_{\perp,k}^2)}, 
\;u_3=\sqrt{2+(d-t_{\perp,k}+s_2)^2/(2 t_{\perp,k}^2)}, 
\;u_4=\sqrt{2+(d-t_{\perp,k}-s_2)^2/(2 t_{\perp,k}^2)}$.

So far the formalism was quite generic. Specializing to Coulomb repulsion
generated positive definite pairing potentials \cite{gk1} for the
cuprates, we consider the realistic 
non-separable form (and thus harder computationally)
\be
V_i(\vec{k},\vec{p})=V_i(\vec{k}-\vec{p}) \;, \;\; V_i(\vec q)=V_{oi} 
\sin^2(q_x a) \sin^2(q_y a)  \;,\;\; V_{oi}>0 \;\; ,
\ee
which is peaked at (near) $\vec Q =(\pm \pi,\pm \pi)$ for $a=0.5(0.5<a<0.6)$.
This type of potential is well known to generate a $d_{x^2-y^2}$-wave
gap \cite{scala,gk1}. 

Another relevant issue is the doping imbalance for M$>$2 layers in the unit 
cell. Namely, NMR experiments show that the outer layers are overdoped 
with holes compared to the
inner layers \cite{nmr}, in agreement with earlier theoretical estimates
\cite{sta}.
We account for this fact by typically considering 
$\mu_{out}=\mu - 0.06 t,\; \mu_{in}=\mu +0.06 t$ for both M=3,4, with 
$\mu_{out}$/$\mu_{in}$ 
referring to the chemical potentials of the original outer/inner layers
and $\mu$ the chemical potential for the case M=1,2.
Assuming a screened electronic interaction, its strength is determined by 
the susceptibility $\chi(q,\om)$ \cite{gk1,gk2}. It can be shown 
that $\chi(q,\om=0)$ is a decreasing function of $\mu$ for the range 
of doping considered herein \cite{gk1}.
Therefore, taking $V_{o1}=V_{o,out}>V_{o2}=V_{o,in}$, reflects also the
effect of $\mu_{out} < \mu_{in}$ for M=3,4. In our model, this potential
imbalance is more important than the sheer $\mu$ imbalance. 
Overall,  
$V_{i}(k,p)$ and $T_k$ (c.f. below)
mostly determine the gaps $\Delta_{i,k}$ and hence $T_c$.
This simply reflects the fact that both $V_{i}(k,p)$ and $T_k$ enter on an 
equal basis in the BCS gap equations - c.f. eqs. (9) and (11). The potential
has a drastic influence on $T_c$, as is known from standard BCS-Eliashberg
theory \cite{scala,gk1}.

We give a summary of a few calculations in the 3 figures and the table below, 
noting that they refer
to $d_{x^2-y^2}$-wave solutions of the gap equations above \cite{sol}. 
In all cases 
the energy scale is given by $t=1$, and $a=0.5$. 
Typically values $0.5<a<0.6$ yield higher $T_c$'s, without a
qualitative change of the results below.

\begin{table}\centering
%\begin{tabular}  

{\bf A. $t'=-0.35, t''=0, n=0.80, T_o=-0.002$ } 

\begin{tabular}{|c|c|c|c|} \hline   \hline

1. & $t_o=0.001$ & $V_{o1}=V_{o2}=4$ & 
$T_c(M=1-4)=(5.384,6.347,{\bf *6.424},6.241) 10^{-2}$  \\

2. & $t_o=0.01$ & $V_{o1}=V_{o2}=4$ &
$T_c(M=1-4)=(5.384,{\bf *6.347},6.187,5.852) 10^{-2}$  \\

3. & $t_o=0.001$ & $V_{o1}=V_{o2}=3$ & 
$T_c(M=1-4)=(1.960,{\bf *2.943},2.806,2.611) 10^{-2}$ \\

3.a  & $t_o=0.001$ & $V_{o1}=3.2, V_{o2}=3$ & 
$T_c(M=3,4)=({\bf *3.403},3.240) 10^{-2}$ \\

3.b & $t_o=0.001$ & $V_{o1}=3.5, V_{o2}=3$ & 
$T_c(M=3,4)=({\bf *4.406},4.282) 10^{-2}$ \\
\hline
\end{tabular}

{\bf B. $t'=-0.35, t''=0, n=0.80, T_o=-0.02$ }     

\begin{tabular}{|c|c|c|c|} \hline  \hline
1. & $t_o=0.001$ & $V_{o1}=4.5, V_{o2}=4$ & 
$T_c(M=1-4)=(5.384,{\bf *14.68},13.02,9.870) 10^{-2}$    \\

\hline
\end{tabular}

{\bf C. $t'=-0.3, t''=0.2, n=0.85, T_o=-0.002$  }

\begin{tabular}{|c|c|c|c|} \hline  \hline
1. & $t_o=0.001$ & $V_{o1}=V_{o2}=4$ & 
$T_c(M=1-4)=(1.215,{\bf *1.700},1.595,1.546) 10^{-2}$ \\

1.a. & $t_o=0.001$ & $V_{o1}=4.5, V_{o2}=4$ & 
$T_c(M=3,4)=({\bf *2.264},2.216) 10^{-2}$ \\

2.& $t_o=0.01$ & $V_{o1}=V_{o2}=4$ & 
$T_c(M=1-4)=(1.215,{\bf *1.632},1.382,0.9957) 10^{-2}$ \\

2.a. & $t_o=0.01$ & $V_{o1}=4.5, V_{o2}=4$ & 
$T_c(M=3,4)=({\bf *1.978},1.627) 10^{-2}$ \\
\hline
\end{tabular}

\caption{ $T_c$ calculated as a function of the parameters shown. }
\end{table}

In the table, an asterisk marks the highest $T_c$ in each case. 
A maximum $T_c$ for M=3, without 
resorting to the potential imbalance $V_{o,out}>V_{o,in}$ ($V_1>V_2$ etc.), 
can be obtained
for a small enough (in magnitude) pair hopping $T_o<0$, and this fact is
facilitated by smaller values of $t_o$ - c.f. case A.1 and the
figures.
Otherwise, for $V_{o,out}=V_{o,in}$ and higher $T_o$, M=2 yields the
maximum $T_c$.
$T_c$ depends weakly on the single particle hopping $t_o$.
In ref. \cite{liec} it was shown 
that the bare value of $t_o$ is significantly reduced through the effect
of interactions.
The sign of $t_o$ is irrelevant, as can be seen from the gap equations.
This is not the case for the sign of the pair-coherent term $T_o$ though
(c.f. below). 
$T_o>0$, {\em in general}, does not 
reproduce the dome-type dependence of $T_c$ on M with a maximum for M=3.

Given the strong dependence of $T_c$ on $T_k$ within our model,
materials with Y, Ba and Sr intercalants should have a much smaller  
$T_k$ than Ca-intercalated cuprates. Moreover, the in-plane pairing potential 
is expected to be weaker for the former materials for M$>$1. 
 
In all, we note that the experimental situation of the $T_c$ increasing
for M=1-3 (or 4) and then dropping, is only realized for 
rather {\em small negative} 
values of the pair tunneling amplitude $T_o$. E.g. for $T_o=-0.02 t$ M=2
yields the maximum $T_c$, as opposed to M=3 for $T_o=-0.002 t$ - c.f.
case D.3 below for a maximum for M=4. The negative $T_k$ can be understood
as follows. Within the frame of second order perturbation theory, 
we have the relation 
$T_{il,k} = -2 t_{\perp,k}^2 \; 
v_{i,k}^{(0)}u_{i,k}^{(0)}v_{l,k}^{(0)}u_{l,k}^{(0)}
/ (E_{i,k}^{(0)}+E_{l,k}^{(0)})$, with the index $(0)$ denoting the initial
{\em uncoupled} layers, 
with $E_{i,k}^{(0)} = \sqrt{(\Delta_{i,k}^{(0)})^2+\varepsilon_{i,k}^2}$. 
$T_{il,k}<0$ follows from $E_{i,k}^{(0)}+E_{l,k}^{(0)}>0$.

We give the results of our calculations for five homologuous series of
cuprates, which are similar to the cases above. We use a restricted
set of parameters - $t=220-245$ meV, $t'=-0.35t$, 
$t''=0$, $t_o=0.03t$, $n=0.80$, which corresponds to optimal doping, 
$a=0.5$, $T_o=-0.0044t - (-0.0018)t$, and 
$V_{oi}=2.7t-4.2t$ - to demonstrate the
fitting capacity of our model. We obtain exact matches with the experimental
$T_c$ values, given in degrees K and taken from ref. \cite{lin}. 
We also give the ratios 
R=max$\{\Delta_{i,k}(T=0)\}/T_c$, which turn out to be in the range 0.71-1.
These values are too small compared to the actual experimental ones, pointing
to the limitations of the BCS description for the cuprates. Along the same 
line, the  $V_{oi}$ values are in the intermediate coupling regime,
i.e. beyond the strict limit of applicability of the BCS framework. 
Note that the opposite potential imbalance $V_1<V_2$ would yield the same 
$T_c$'s with somewhat higher values of $V_2$ for M=3,4 than the $V_1$ 
values below. The values of $t$ and $V_{o1}$ are adjusted so as
to yield $T_c$(M=1). Then, $T_o$ is chosen so as to yield $T_c$(M=2).
Subsequently, $V_{o2}(M=3,4)=V_{o1}(M=1)=V_{o1}(M=2)$, and $V_{o1}(M=3,4)$ 
are chosen so as to yield $T_c$(M=3,4).

{\bf D.1} Bi$_2$Sr$_2$Ca$_{M-1}$Cu$_M$O$_{2M+4+\delta}$ with $t=245$ meV and 
$T_o=-0.0044t$, $V_{o1}=V_{o2}=2.7t$.
For M=3 we take $V_{o1}=3.5t$, thus
obtaining $T_c$(M=1-3)=36, 90, 110 K and 
R(M=1-3)=0.769, 0.713, 0.867.

{\bf D.2} TlBa$_2$Ca$_{M-1}$Cu$_M$O$_{2M+3+\delta}$ with $t=229$ meV and 
$T_o=-0.0044t$,
$V_{o1}=V_{o2}=3t$.
For M=3 we take $V_{o1}=3.85t$ and for M=4 $V_{o1}=4.16t$, thus
obtaining $T_c$(M=1-4)=52, 107, 133, 127 K and
R(M=1-4)=0.765, 0.781, 0.973, 1.034.

{\bf D.3} Tl$_2$Ba$_2$Ca$_{M-1}$Cu$_M$O$_{2M+3+\delta}$ with $t=220$ meV and 
$T_o=-0.0024 t$,
$V_{o1}=V_{o2}=3t$.
For M=3 we take $V_{o1}=3.67t$ and for M=4 $V_{o1}=4.2t$, thus
obtaining $T_c$(M=1-4)=50, 80, 110, 122 K and
R(M=1-4)=0.765, 0.755, 0.909, 1.039. This series has maximum $T_c$
for M=4 - c.f. \cite{burn}. 

{\bf D.4} Tl$_2$Ba$_2$Ca$_{M-1}$Cu$_M$O$_{2M+4+\delta}$ with $t=221$ meV and 
$T_o=-0.0018t$, $V_{o1}=V_{o2}=3.5t$.
For M=3 we take $V_{o1}=3.76t$ and for M=4 $V_{o1}=3.83t$, thus
obtaining $T_c$(M=1-4)=90, 115, 125, 116 K and
R(M=1-4)=0.795, 0.845, 0.934, 0.976.

{\bf D.5} HgBa$_2$Ca$_{M-1}$Cu$_M$O$_{2M+2+\delta}$ with $t=238.5$ meV and 
$T_o=-0.0021t$, $V_{o1}=V_{o2}=3.5t$.
For M=3 we take $V_{o1}=3.75t$ and for M=4 $V_{o1}=3.89t$, thus
obtaining $T_c$(M=1-4)=97, 127, 135, 129 K and
R(M=1-4)=0.765, 0.851, 0.932, 0.979.

\vspace{.3cm}
In summary, higher $T_c$ in multilayer copper oxides 
arises from a relatively increased 
(in magnitude) pair interlayer coupling $T_k < 0$, combined 
with a substantial strength of the repulsive Coulomb interaction between
the electrons. It seems 
that all these ingredients are present in the Ca-intercalated materials.
This picture should possibly be complemented by the CuO$_2$ lattice
symmetry effects \cite{gk3}. The BCS approach,
though of limited applicability in the cuprates, offers relevant insight.
We give an interpretation of the negative $T_k$ in the BCS framework. 

\vspace{.3cm}
Stamatis Stamatiadis provided essential programming assistance.
The author enjoyed discussions with Josheph Betouras, who also read
critically the manuscript.
Funding received from the Dept. of Materials 
Science of the University of Crete is acknowledged.

\vspace{.3cm}
$^*$ E-mail address : kast@iesl.forth.gr, giwkast@gmail.com

\begin{figure}[ht]
\includegraphics[height=7.5cm]{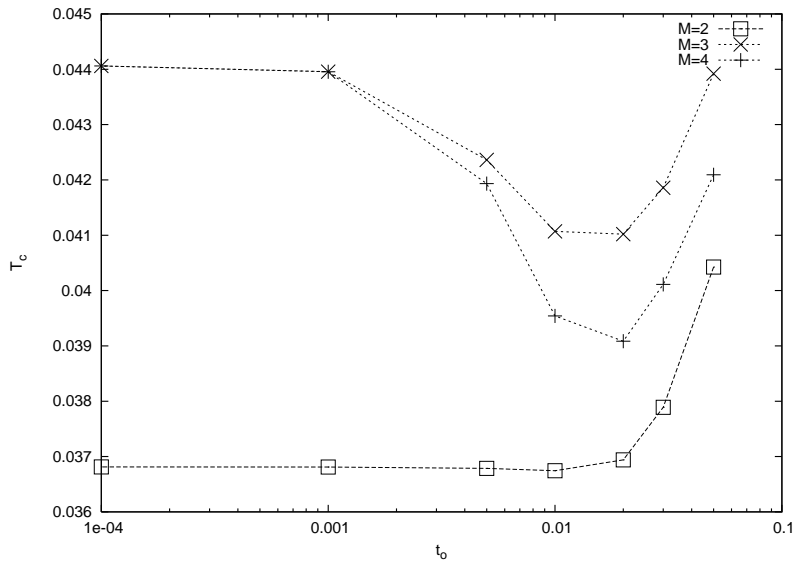}
\caption{$T_c$ (in units of $t$) as a function of interplane hopping for
$n=0.85$, $t'=-0.35$, $t''=0$, $V_{o1}=V_{o2}=4$. The inner/outer layer
doping imbalance is taken into account for M=3,4. Squares : M=2,
x's : M=3 and crosses : M=4. $T_c$ versus $t_o$ is shown for
$T_o=0$. All lines are guides to the eye.}
\end{figure}

\begin{figure}[ht]
\includegraphics[height=7.5cm]{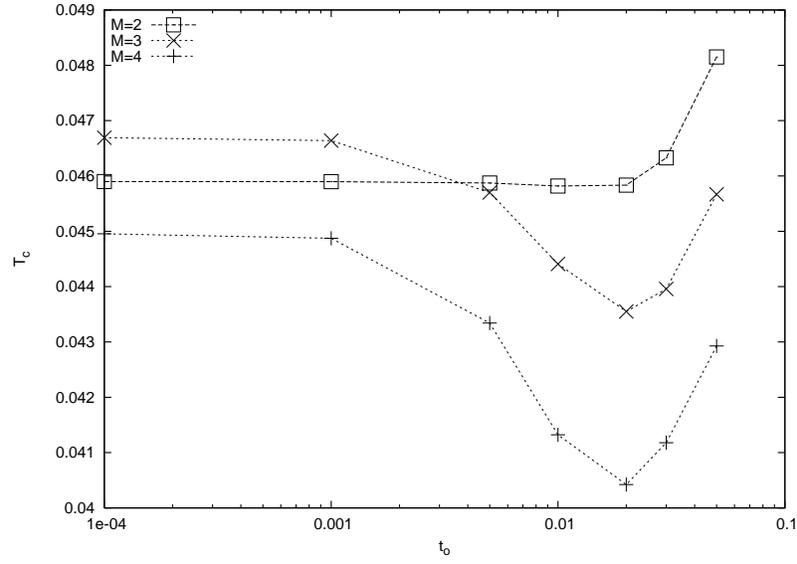}
\caption{ Same as in fig. 1. $T_c$ versus $t_o$ for $T_o=-0.002$ .}
\end{figure}

\begin{figure}[ht]
\includegraphics[height=7.5cm]{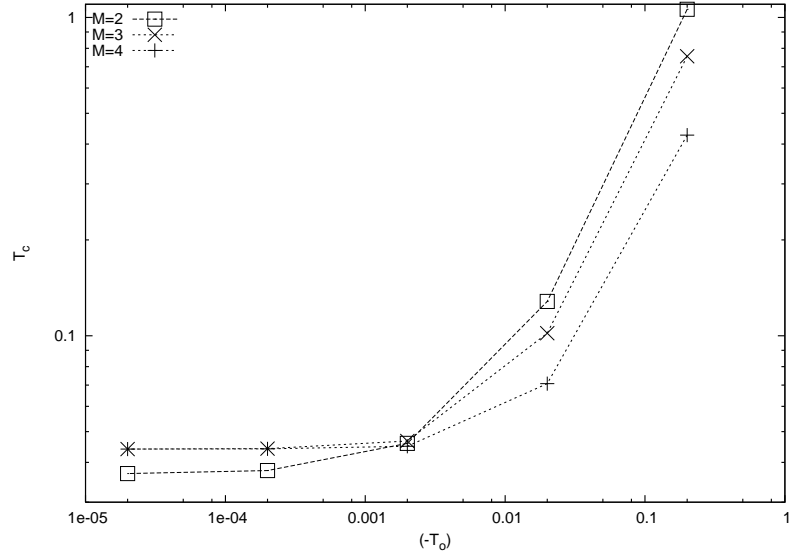}
\caption{ Same as in fig. 1. $T_c$ versus (-$T_o$) for $t_o=10^{-7}$. 
Notice the log scale for $T_c$. }
\end{figure}


\begin{thebibliography}{99}


\bibitem{lin} X.J. Chen and H.Q. Lin, Phys. Rev. B {\bf 69}, 104518 (2004),
and Phys. Rev. B {\bf 71}, 109901(E) (2005).

\bibitem{burn}
G. Burns, High-Temperature Superconductivity: An Introduction (Academic,
Boston, 1992).

\bibitem{le} A. J. Leggett, J. Phys. Chem. Solids {\bf 59}, 1729 (1998).

\bibitem{tes} Z. Tesanovic, Phys. Rev. B {\bf 36}, 2364 (1987).

\bibitem{sta} M. Di Stasio, K.A. M\"uller, and L. Pietronero, 
Phys. Rev. Lett. {\bf 64}, 2827 (1990).

\bibitem{ch} S. Chakravarty, A. Sudbo, P.W. Anderson, and S. Strong, 
Science {\bf 261}, 337 (1993).

\bibitem{spa} K. Byczuk and J. Spalek, Phys. Rev. B {\bf 53}, R518 (1996).

\bibitem{liec} A. I. Liechtenstein, O. Gunnarsson, O. K. Andersen, 
and R. M. Martin, Phys. Rev. B {\bf 54}, 12505 (1996).

\bibitem{car} E. W. Carlson, S. A. Kivelson, V. J. Emery, and E. Manousakis 
Phys. Rev. Lett. {\bf 83}, 612 (1999).

\bibitem{cha} S. Chakravarty, Hae-Young Kee, and K. V\"olker,
Nature {\bf 428}, 53 (2004).

%\bibitem{}

\bibitem{scala} D.J. Scalapino, Phys. Repts. {\bf 250}, 329 (1995).

\bibitem{gk1} G. Kastrinakis, Physica C {\bf 340}, 119 (2000). 
%C.f. eqs. (29)-(32) for the pairing potential.

\bibitem{gk2} G. Kastrinakis, Phys. Rev. B. {\bf 71}, 014520 (2005).

\bibitem{dg} P.G. de Gennes, Superconductivity of Metals and Alloys 
(Benjamin, New York, 1966).

\bibitem{lie2} O.K. Andersen, A.I. Liechtenstein, O. Jepsen, and F. Paulsen,
J. Phys. Chem. Solids {\bf 56}, 1573 (1995).

\bibitem{oper}
For M=3,4
\be
U_3 =
\left( \begin{array}{ccc}
-b_1(k) & b_2(k) & b_3(k) \\
g_1(k) & -g_2(k) & g_3(k) \\		 
b_1(k) & b_2(k) & b_3(k) 
\end{array} \right)
\;\;, \;\;
U_4 = \left(\begin{array}{cccc}
b_1(k) & -b_2(k) & b_3(k) & b_4(k)\\
g_1(k) & -g_2(k) & g_3(k) & g_4(k)\\		 
-g_1(k) & g_2(k) & g_3(k) & g_4(k)\\		
-b_1(k) & b_2(k) & b_3(k) & b_4(k)\\
\end{array} 
\right) \;\;,
\ee
with $b_i(k),g_i(k)$ as given after eqs. (\ref{exi3}) and (\ref{exi4})
respectively.

\bibitem{nmr} H. Kotegawa, Y. Tokunaga, K. Ishida, G.-q. Zheng, Y. Kitaoka, 
K. Asayama, H. Kito, A. Iyo, H. Ihara, K. Tanaka, K. Tokiwa, and 
T. Watanabe, J. Phys. Chem. Solids {\bf 62}, 171 (2001).

\bibitem{sol}
We use a 64 by 64
discretization of the Brillouin zone, and we solve the full temperature 
dependent gap equations 
directly for the angles $\theta_{i,k}$. 

\bibitem{gk3} G. Kastrinakis, e-print arxiv:0901.0097 .

\end{thebibliography}
\end{document}